\RequirePackage{lineno}
\documentclass[aps,prl,twocolumn,superscriptaddress,showpacs,preprintnumbers,amsmath,amssymb, nobibnotes]{revtex4-1}

\usepackage{graphicx} % Include figure files
\usepackage{dcolumn}  % Align table columns on decimal point
\graphicspath{{ps}}
\usepackage{enumitem}  % Align table columns on decimal point
\usepackage{feynmf}    % for feynmandiagram  
\usepackage{xcolor}
\newcommand{\bks}{$B^{0} \to K_S^{0}K_S^{0}\gamma$}

\begin{document}

%\vspace*{-3\baselineskip}
%\resizebox{!}{3cm}{\includegraphics{belle.eps}}

\preprint{\vbox{ \hbox{   }
						\hbox{Belle preprint {\it 2022-03}}
            \hbox{KEK preprint {\it 2021-62}}
  		      % \hbox{hep-ex nnnn}
}}

\title{ \quad\\[1.0cm] Search for \bks ~decays at Belle}

%%%% >>>>> insert the authorlist here. BEFORE the abstract !!!!! <<<<<
%%%% >>>>> from the authorship confirmation web page
%%% Name the file author.tex and use \input{author} to insert into your latex file.
 %%% Paper:    B0 to KS KS gamma
%%% Journal:  Physical Review D
%%% Contacts: K.H. Kang (kookhyun.kang@ipmu.jp)
%%%           H.B. Park (sunshine@knu.ac.kr)
%%%           H.B. Jeon (hyebin@uchicago.edu)
%%%           A. Ishikawa (akimasa.ishikawa@kek.jp)
%%% Non-responding authors or those who said NO are commented out.
%%% ====================================================================
%%% Click the RELOAD button on your web browser to see the updated file.
%%% ====================================================================
%%% Use \input{author} to insert this material into your latex file.
%%%%% Force institutions to appear in alphabetical order when typeset.
\noaffiliation
\affiliation{Department of Physics, University of the Basque Country UPV/EHU, 48080 Bilbao}
\affiliation{University of Bonn, 53115 Bonn}
\affiliation{Brookhaven National Laboratory, Upton, New York 11973}
\affiliation{Budker Institute of Nuclear Physics SB RAS, Novosibirsk 630090}
\affiliation{Faculty of Mathematics and Physics, Charles University, 121 16 Prague}
%%%\affiliation{Chiba University, Chiba 263-8522}
\affiliation{Chonnam National University, Gwangju 61186}
\affiliation{Chung-Ang University, Seoul 06974}
\affiliation{University of Cincinnati, Cincinnati, Ohio 45221}
\affiliation{Deutsches Elektronen--Synchrotron, 22607 Hamburg}
\affiliation{Duke University, Durham, North Carolina 27708}
\affiliation{Institute of Theoretical and Applied Research (ITAR), Duy Tan University, Hanoi 100000}
\affiliation{University of Florida, Gainesville, Florida 32611}
\affiliation{Department of Physics, Fu Jen Catholic University, Taipei 24205}
\affiliation{Key Laboratory of Nuclear Physics and Ion-beam Application (MOE) and Institute of Modern Physics, Fudan University, Shanghai 200443}
\affiliation{Justus-Liebig-Universit\"at Gie\ss{}en, 35392 Gie\ss{}en}
\affiliation{Gifu University, Gifu 501-1193}
%%%\affiliation{II. Physikalisches Institut, Georg-August-Universit\"at G\"ottingen, 37073 G\"ottingen}
\affiliation{SOKENDAI (The Graduate University for Advanced Studies), Hayama 240-0193}
\affiliation{Gyeongsang National University, Jinju 52828}
\affiliation{Department of Physics and Institute of Natural Sciences, Hanyang University, Seoul 04763}
\affiliation{University of Hawaii, Honolulu, Hawaii 96822}
\affiliation{High Energy Accelerator Research Organization (KEK), Tsukuba 305-0801}
\affiliation{J-PARC Branch, KEK Theory Center, High Energy Accelerator Research Organization (KEK), Tsukuba 305-0801}
\affiliation{National Research University Higher School of Economics, Moscow 101000}
\affiliation{Forschungszentrum J\"{u}lich, 52425 J\"{u}lich}
%%%\affiliation{Hiroshima University, Higashi-Hiroshima, Hiroshima 739-8530}
\affiliation{IKERBASQUE, Basque Foundation for Science, 48013 Bilbao}
\affiliation{Indian Institute of Science Education and Research Mohali, SAS Nagar, 140306}
\affiliation{Indian Institute of Technology Bhubaneswar, Satya Nagar 751007}
\affiliation{Indian Institute of Technology Guwahati, Assam 781039}
\affiliation{Indian Institute of Technology Hyderabad, Telangana 502285}
\affiliation{Indian Institute of Technology Madras, Chennai 600036}
\affiliation{Indiana University, Bloomington, Indiana 47408}
\affiliation{Institute of High Energy Physics, Chinese Academy of Sciences, Beijing 100049}
\affiliation{Institute of High Energy Physics, Vienna 1050}
\affiliation{Institute for High Energy Physics, Protvino 142281}
%%%\affiliation{Institute of Mathematical Sciences, Chennai 600113}
\affiliation{INFN - Sezione di Napoli, I-80126 Napoli}
\affiliation{INFN - Sezione di Roma Tre, I-00146 Roma}
\affiliation{INFN - Sezione di Torino, I-10125 Torino}
\affiliation{Iowa State University, Ames, Iowa 50011}
\affiliation{Advanced Science Research Center, Japan Atomic Energy Agency, Naka 319-1195}
\affiliation{J. Stefan Institute, 1000 Ljubljana}
%%%\affiliation{Kanagawa University, Yokohama 221-8686}
\affiliation{Institut f\"ur Experimentelle Teilchenphysik, Karlsruher Institut f\"ur Technologie, 76131 Karlsruhe}
\affiliation{Kavli Institute for the Physics and Mathematics of the Universe (WPI), University of Tokyo, Kashiwa 277-8583}
%%%\affiliation{King Abdulaziz City for Science and Technology, Riyadh 11442}
\affiliation{Department of Physics, Faculty of Science, King Abdulaziz University, Jeddah 21589}
\affiliation{Kitasato University, Sagamihara 252-0373}
\affiliation{Korea Institute of Science and Technology Information, Daejeon 34141}
\affiliation{Korea University, Seoul 02841}
%%%\affiliation{Kyoto Sangyo University, Kyoto 603-8555}
\affiliation{Kyungpook National University, Daegu 41566}
%%%\affiliation{Universit\'{e} Paris-Saclay, CNRS/IN2P3, IJCLab, 91405 Orsay}
%%%\affiliation{\'Ecole Polytechnique F\'ed\'erale de Lausanne (EPFL), Lausanne 1015}
\affiliation{P.N. Lebedev Physical Institute of the Russian Academy of Sciences, Moscow 119991}
\affiliation{Liaoning Normal University, Dalian 116029}
\affiliation{Faculty of Mathematics and Physics, University of Ljubljana, 1000 Ljubljana}
\affiliation{Ludwig Maximilians University, 80539 Munich}
\affiliation{Luther College, Decorah, Iowa 52101}
\affiliation{Malaviya National Institute of Technology Jaipur, Jaipur 302017}
%%%\affiliation{University of Malaya, 50603 Kuala Lumpur}
\affiliation{Faculty of Chemistry and Chemical Engineering, University of Maribor, 2000 Maribor}
\affiliation{Max-Planck-Institut f\"ur Physik, 80805 M\"unchen}
\affiliation{School of Physics, University of Melbourne, Victoria 3010}
\affiliation{University of Mississippi, University, Mississippi 38677}
\affiliation{University of Miyazaki, Miyazaki 889-2192}
\affiliation{Moscow Physical Engineering Institute, Moscow 115409}
\affiliation{Graduate School of Science, Nagoya University, Nagoya 464-8602}
%%%\affiliation{Kobayashi-Maskawa Institute, Nagoya University, Nagoya 464-8602}
\affiliation{Universit\`{a} di Napoli Federico II, I-80126 Napoli}
\affiliation{Nara Women's University, Nara 630-8506}
\affiliation{National Central University, Chung-li 32054}
\affiliation{National United University, Miao Li 36003}
\affiliation{Department of Physics, National Taiwan University, Taipei 10617}
\affiliation{H. Niewodniczanski Institute of Nuclear Physics, Krakow 31-342}
\affiliation{Nippon Dental University, Niigata 951-8580}
\affiliation{Niigata University, Niigata 950-2181}
%%%\affiliation{University of Nova Gorica, 5000 Nova Gorica}
\affiliation{Novosibirsk State University, Novosibirsk 630090}
\affiliation{Okinawa Institute of Science and Technology, Okinawa 904-0495}
\affiliation{Osaka City University, Osaka 558-8585}
\affiliation{Pacific Northwest National Laboratory, Richland, Washington 99352}
\affiliation{Panjab University, Chandigarh 160014}
%%%\affiliation{Peking University, Beijing 100871}
\affiliation{University of Pittsburgh, Pittsburgh, Pennsylvania 15260}
%%%\affiliation{Punjab Agricultural University, Ludhiana 141004}
%%%\affiliation{Research Center for Electron Photon Science, Tohoku University, Sendai 980-8578}
\affiliation{Research Center for Nuclear Physics, Osaka University, Osaka 567-0047}
\affiliation{Meson Science Laboratory, Cluster for Pioneering Research, RIKEN, Saitama 351-0198}
%%%\affiliation{Theoretical Research Division, Nishina Center, RIKEN, Saitama 351-0198}
%%%\affiliation{RIKEN BNL Research Center, Upton, New York 11973}
\affiliation{Dipartimento di Matematica e Fisica, Universit\`{a} di Roma Tre, I-00146 Roma}
\affiliation{Department of Modern Physics and State Key Laboratory of Particle Detection and Electronics, University of Science and Technology of China, Hefei 230026}
%%%\affiliation{Seoul National University, Seoul 08826}
\affiliation{Showa Pharmaceutical University, Tokyo 194-8543}
%%%\affiliation{Soochow University, Suzhou 215006}
\affiliation{Soongsil University, Seoul 06978}
%%%\affiliation{University of South Carolina, Columbia, South Carolina 29208}
%%%\affiliation{Stefan Meyer Institute for Subatomic Physics, Vienna 1090}
\affiliation{Sungkyunkwan University, Suwon 16419}
\affiliation{School of Physics, University of Sydney, New South Wales 2006}
\affiliation{Department of Physics, Faculty of Science, University of Tabuk, Tabuk 71451}
\affiliation{Tata Institute of Fundamental Research, Mumbai 400005}
%%%\affiliation{Excellence Cluster Universe, Technische Universit\"at M\"unchen, 85748 Garching}
\affiliation{Department of Physics, Technische Universit\"at M\"unchen, 85748 Garching}
\affiliation{School of Physics and Astronomy, Tel Aviv University, Tel Aviv 69978}
\affiliation{Toho University, Funabashi 274-8510}
\affiliation{Department of Physics, Tohoku University, Sendai 980-8578}
\affiliation{Earthquake Research Institute, University of Tokyo, Tokyo 113-0032}
\affiliation{Department of Physics, University of Tokyo, Tokyo 113-0033}
\affiliation{Tokyo Institute of Technology, Tokyo 152-8550}
\affiliation{Tokyo Metropolitan University, Tokyo 192-0397}
%%%\affiliation{Utkal University, Bhubaneswar 751004}
\affiliation{Virginia Polytechnic Institute and State University, Blacksburg, Virginia 24061}
\affiliation{Wayne State University, Detroit, Michigan 48202}
\affiliation{Yamagata University, Yamagata 990-8560}
\affiliation{Yonsei University, Seoul 03722}
  \author{H.~B.~Jeon}\altaffiliation[Now at~]{the Enrico Fermi Institute of the University of Chicago, Chicago,  IL 60637}\affiliation{Kyungpook National University, Daegu 41566} % Kyungpook
  \author{K.~H.~Kang}\affiliation{Kavli Institute for the Physics and Mathematics of the Universe (WPI), University of Tokyo, Kashiwa 277-8583} % IPMU
  \author{H.~Park}\affiliation{Kyungpook National University, Daegu 41566} % Kyungpook
  \author{I.~Adachi}\affiliation{High Energy Accelerator Research Organization (KEK), Tsukuba 305-0801}\affiliation{SOKENDAI (The Graduate University for Advanced Studies), Hayama 240-0193} % KEK
% \author{K.~Adamczyk}\affiliation{H. Niewodniczanski Institute of Nuclear Physics, Krakow 31-342} % Krakow
% \author{J.~K.~Ahn}\affiliation{Korea University, Seoul 02841} % Korea
  \author{H.~Aihara}\affiliation{Department of Physics, University of Tokyo, Tokyo 113-0033} % Tokyo
  \author{S.~Al~Said}\affiliation{Department of Physics, Faculty of Science, University of Tabuk, Tabuk 71451}\affiliation{Department of Physics, Faculty of Science, King Abdulaziz University, Jeddah 21589} % Tabuk
  \author{D.~M.~Asner}\affiliation{Brookhaven National Laboratory, Upton, New York 11973} % BNL
  \author{H.~Atmacan}\affiliation{University of Cincinnati, Cincinnati, Ohio 45221} % Cincinnati
% \author{V.~Aulchenko}\affiliation{Budker Institute of Nuclear Physics SB RAS, Novosibirsk 630090}\affiliation{Novosibirsk State University, Novosibirsk 630090} % BINP
  \author{T.~Aushev}\affiliation{National Research University Higher School of Economics, Moscow 101000} % HSE
  \author{R.~Ayad}\affiliation{Department of Physics, Faculty of Science, University of Tabuk, Tabuk 71451} % Tabuk
% \author{T.~Aziz}\affiliation{Tata Institute of Fundamental Research, Mumbai 400005} % Tata
% ...
% Lebedev
% \author{V.~Zhulanov}\affiliation{Budker Institute of Nuclear Physics SB RAS, Novosibirsk 630090}\affiliation{Novosibirsk State University, Novosibirsk 630090} % BINP
\collaboration{The Belle Collaboration}

%% end author list

\begin{abstract}
We report the first search for the penguin-dominated process $B^0 \rightarrow K_S^0 K_S^0 \gamma$ using the full data sample of $772\times
10^6$ 
$B\bar{B}$ 
pairs collected with the Belle detector at the KEKB asymmetric-energy $e^+ e^-$ collider. 
We do not observe any statistically significant signal yield in the $K_S^0$-pair invariant mass  
range 1 GeV/$c^2 < M_{K_S^0 K_S^0} < $  3 GeV/$c^2$, 
and set the following upper limits 
at 90\% confidence level:
$\mathcal{B}(B^0 \to K_S^0 K_S^0 \gamma) < 5.8\times10^{-7}$, 
$\mathcal{B}(B^0 \to f_2 \gamma)\times \mathcal{B}(f_2 (1270) \to K_S^0 K_S^0 ) < 3.1\times10^{-7}$, and
$\mathcal{B}(B^0 \to f_2^{\prime} \gamma)\times \mathcal{B}(f_2^{\prime} (1525) \to K_S^0 K_S^0 ) < 2.1\times10^{-7}$.
Further, 90\% confidence upper limits 
have also been set
in the range of [0.7--2.9]$\times10^{-7}$ 
on the 
$B^0 \rightarrow K_S^0 K_S^0 \gamma$
branching fraction in bins of 
$M_{K_S^0 K_S^0}$. 
\end{abstract}

\maketitle
%%%% >>>> keep the final version single-spaced
\tighten
%\saythanks
{\renewcommand{\thefootnote}{\fnsymbol{footnote}}}
\setcounter{footnote}{0}

%%%
%%%From here introduction
%%%
Radiative $b\to s \gamma$ and $b\to d \gamma$ quark transitions
are flavor-changing-neutral-current processes
and 
not allowed at the tree level in the Standard Model (SM). 
Such decays proceed predominantly through the radiative loop diagrams
and are sensitive to the contributions from non-SM particles which may enter the loop diagram.
For example, the two-Higgs-doublet model (2HDM) introduces
an additional Higgs doublet and 
the charged Higgs 
may appear in the loop instead of a $W$ boson. 
Wilson coefficients in the operator product expansion~\cite{OPE} are modified to include the effect of
the 2HDM~\cite{EWP} and this new term depends on the mass of the charged Higgs~\cite{THDM}. 
Thus, any disparity in the branching fraction from the SM expectations can be interpreted as a new physics contribution.

Branching fractions of several exclusive $b\to s \gamma$ modes have been measured:
$B \to K^* \gamma$~\cite{KstarGamma}; $B \to K_1 (1270) \gamma$~\cite{K1270Gamma}; $B\to \phi K \gamma$~\cite{PhiKGamma}; 
$B\to K \eta^{\prime}\gamma$~\cite{KEtaPrimeGamma}; $B\to K \eta \gamma$~\cite{KEtaGamma}.
On the other hand, $B\to \rho\gamma$ and $B\to \omega\gamma$ 
are the only observed exclusive $b\to d\gamma$ modes~\cite{rhogamma},
and a further study of an additional 
exclusive mode
is important to constrain the ratio of Cabibbo-Kobayashi-Maskawa matrix elements $|V_{td}/V_{ts}|$ ~\cite{CKM}
and also
to test the theoretical models.

\begin{figure}[!h]
	\centering
		\includegraphics[width=0.5\textwidth]{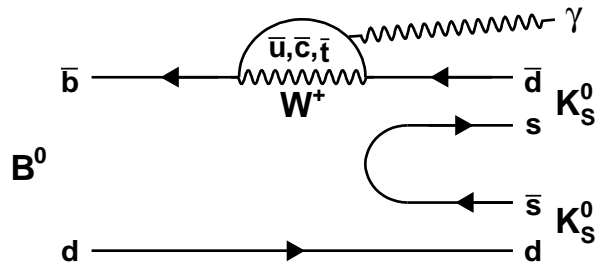} 
	\caption{$b \to d\gamma$ penguin diagram for $B^0 \to K_S^0 K_S^0 \gamma$ decay.}
	\label{fig:feynman}
\end{figure}

The $B^0 \rightarrow K_S^0 K_S^0 \gamma$ decay shown in Fig.~\ref{fig:feynman} is one such radiative electroweak penguin process 
that proceeds via $b \rightarrow d\gamma$ at the quark level.
The angular momentum of an intermediate state decaying into two identical neutral pseudoscalar particles, $K_S^0$, 
is even due to Bose-Einstein statistics.
The spin of the $K_S^0 K_S^0$ system must be at least two by the conservation of angular momentum
in the $B^0 \rightarrow K_S^0 K_S^0 \gamma$ decay,
since the photon is a massless vector particle.
Therefore, spin-2 is the lowest spin state possible for the $K_S^0 K_S^0$ system.
In this Letter, we present results from a search for the $B^0 \to K_S^0 K_S^0 \gamma$ decay.

The $\Upsilon(4S)$ meson is produced at the KEKB asymmetric-energy $e^+ e^-$ collider~\cite{KEKB} 
with electrons and positrons having the energes of 8 GeV and 3.5 GeV, respectively, 
and subsequently decays to $B\bar{B}$ pairs which are nearly at rest in the center-of-mass system (CMS).
The $z$ axis is defined opposite to the $e^+$ beam direction.
We search for the decay $B^0 \rightarrow K_S^0 K_S^0 \gamma$ using the full data sample of 
$(772 \pm 11)\times 10^{6} ~B\bar{B}$ pairs collected at the $\Upsilon(4S)$  resonance with the Belle detector 
at the KEKB asymmetric-energy $e^+ e^-$ collider. 
This is the first ever search for a $B^0$ decay to two pseudoscalars $K_S^0$ with a prompt photon in the final state.
The inclusion of the charge conjugate modes is implied throughout this paper unless otherwise stated.

%%%
%%%From here Detector 
%%%
The Belle detector is a hermetic magnetic spectrometer to detect the decay products of $B$ mesons 
that consists of a silicon vertex detector, a 50-layer central drift chamber (CDC), 
an array of aerogel threshold Cherenkov counters, a barrel-like arrangement of time-of-flight scintillation counters, 
and an electromagnetic calorimeter comprised of CsI(Tl) scintillation crystals (ECL). 
These detector components, providing high vertex resolution, good tracking, sophisticated particle identification capability, 
and excellent calorimetry, are located inside a superconducting solenoid coil providing a 1.5 T magnetic field.
An iron flux-return is located outside of the magnetic coil which is instrumented to detect $K_L^0$ mesons and identify muons.
The detector is described in detail elsewhere~\cite{BELLE}.

%%%
%%%From here MCintroduction 
%%%
The event selections are optimized using simulated Monte Carlo (MC) samples.
The MC samples for the signal and the backgrounds are generated with E{\scriptsize VT}G{\scriptsize EN}~\cite{Evtgen} 
and the detector response is then simulated using G{\scriptsize EANT}3~\cite{GEANT3}. 
Any environmental changes in the Belle detector and KEKB accelerator machine during the operations are reflected in the detector simulation.
To generate the signal MC sample of 
$B^0$ decaying to a tensor meson (as an intermediate state) and a prompt photon, 
a two-body decay model is used with appropriate helicity amplitudes. 
Then, the intermediate state decays to two $K_S^0$. 
As the $K_S^0 K_S^0$ structure in $B^0 \rightarrow K_S^0 K_S^0 \gamma$ is unknown,
the mass of the intermediate state is evenly distributed between 1 GeV/$c^2$ and 3 GeV/$c^2$ in the signal MC sample.

%%%
%%%From here Selection 
%%%
Photons must have no associated tracks in the CDC, be in the ECL barrel region 
($33^\circ < \theta_{\gamma} < 128^\circ$), and have
a ratio of energy deposition in 3$\times$3 ECL crystals to that in 5$\times$5 ECL crystals 
centered on the crystal having the maximum energy 
above 95\%. 
The CM energy of the prompt photon candidate, $E_{\gamma}$, must satisfy the requirement 1.6 GeV $< E_{\gamma} <$ 2.8 GeV. 
Most background photons originate from $\pi^0$ and $\eta$ to $\gamma\gamma$ decays. 
We combine the photon candidate with all other photons with momenta larger than
50 MeV/$c$
in the event and 
calculate the probabilities of the reconstructed particle to be $\pi^0$-like or $\eta$-like~\cite{Koppenburg}.
The background photons are suppressed by removing $\pi^0$-like and
$\eta$-like photons by applying selection criteria on a likelihood based selector. 
About 86\% of the photons from the signal $B$ are retained and about 62\%
from the accompanying $B$ are rejected.
For  more than one photon satisfying the selection criteria for the prompt photon candidate, 
the most energetic photon is chosen as the prompt photon candidate.
The selection efficiency of the prompt photon is approximately 50\% and
99.5\% are truly matched photons for the signal MC.

% Ks selection
$K_S^0$ candidates are reconstructed from two oppositely charged tracks.
A displaced vertex consistent with $K_S^0 \rightarrow \pi^+ \pi^-$ decay is required 
using a neural network (NN) discriminator with 20 inputs~\cite{KSelection}; 
this selection also suppresses
$\Lambda \rightarrow p \pi^-$ decays. 
The invariant mass of the pion pairs is then required to satisfy $|M_{\pi\pi} - m_{K_S^0}| < 4.7$ MeV/$c^2$, 
corresponding to a $\pm 2.6\sigma$ interval in mass resolution, 
where $m_{K_S^0}$ is the nominal $K_S^0$ mass~\cite{PDG2020}.
$B^0$ candidates are formed by combining two $K_S^0$ candidates and one prompt photon candidate. 
The energy difference $\Delta E \equiv E_{B}^{\rm cms} - E_{\rm beam}^{\rm cms}$ and 
the beam-energy-constrained mass ${M_{\rm bc} \equiv \sqrt{(E_{\rm beam}^{\rm cms})^2 - |\vec{p}_B^{\rm\;cms}|^{2} c^{2}}/c^2}$, 
where $E_{\rm beam}^{\rm cms}$ is the beam energy, and $E_{B}^{\rm cms}$ and $\vec{p}_{B}^{\rm cms}$ are the energy and momentum 
of the reconstructed $B^0$, respectively, are used to identify $B^0$ candidates. 
The candidates satisfying the requirements $5.20~ {\rm GeV/}c^2 < M_{\rm bc} < 5.29~ {\rm GeV/}c^2$ and $|\Delta E | < 0.5$ GeV are retained for further analysis. 
We find that 6\% of the events have more than one $B^0$ candidate.
In case of multiple candidates, we choose the one with the smallest $\chi^2$ as defined by
$\chi^2 = \sum_{i=1}^{2}  [(m_{K_S^0} - M_i(\pi^{+}\pi^{-})) / \sigma_{\pi\pi}]^2$, 
where $\sigma_{\pi \pi}$ is the mass resolution for the reconstructed $K^0_S$. 

%%%
%%%From here Background 
%%%
The dominant background is $e^+ e^- \rightarrow q\bar{q} ~(q= u, d, s, c)$ continuum events. 
We use another NN with four input variables calculated in the CMS 
to suppress this background~\cite{NB}: 
the cosine of the polar angle ($\cos \theta_B$) of the $B^0$ candidate flight direction; 
the cosine of the angle ($\cos \theta_T$)
between the thrust axis of the $B^0$ candidate and that of the rest of the event; 
a flavor-tagging quality parameter of the accompanying $B$ meson~\cite{Hamlet}; 
and a likelihood ratio obtained from the modified Fox-Wolfram moments~\cite{KSFW}. 
The NN outputs for the signal and continuum MC events peak at +1 and –1, respectively. 
The figure-of-merit (FOM) is calculated as ~\cite{FOMeq}:

\begin{equation}
        \text{FOM} = \frac{\epsilon_S(t)}{a/2+\sqrt{N_{\text{bkg}}(t)}},
        \label{eq:punzi}
\end{equation}

\noindent
where $t$ is the NN output; $\epsilon_S(t)$~is the signal efficiency as a function of $t$ determined by using the signal MC sample; 
$N_{\text{bkg}}$ is the remaining background events after NN selection 
and $a$ is taken to be 3 for a
3$\sigma$ significance due to the low signal-to-background ratio, as suggested in Ref.~\cite{FOMeq}.
The maximal FOM is obtained at 0.93 which rejects 99\% of the continuum MC events and retains 37\% of the signal MC events.
Since we expect a few signal events and relatively large backgrounds, 
we further suppress the continuum background by using the helicity angle, $\theta_H$,  
which is the angle between the direction opposite to the $B^0$ candidate and 
that of the $K_S^0$ momentum in the rest frame of the $K_S^0 K_S^0$ system. 
To maximize the FOM, we require $0.24  < |\cos \theta_H | < 0.86$ which removes 60\% of the background while retaining 86\% of the signal.

We use a Crystal Ball line shape~\cite{Crystal} and a first-order polynomial for the signal and mis-reconstructed components, respectively. 
The signal region is defined as 
$-0.16~ {\rm GeV} < \Delta E < 0.09~ {\rm GeV}$ and 
$5.272~ {\rm GeV/}c^2 < M_{\rm bc} < 5.290~ {\rm GeV/}c^2$,
corresponding to $\pm 3 \sigma$ windows.
About 99\% of $B^0$ candidates in the signal region correctly match signal $B^0$ and
all of them have the correct prompt photon as the daugther particle of the signal $B^0$.

%(2) Generic B decay background
From continuum MC, we estimate 2.2 events in the signal region.
In addition to the continuum, various $B\bar{B}$ background sources are also
studied. 
Both neutral and charged $B\bar{B}$ MC samples corresponding to an integrated luminosity six times 
larger than that of the full data sample are used. 
We expect 0.3 events from generic $B\bar B$ decays in the signal region.
The decay $B^0 \to D^0 (\to K_S^0 \pi^0 ) K^0 $, with a branching fraction of $5.2\times10^{-5}$
~\cite{PDG2020}, 
is treated separately from the generic $B\bar B$ because its $\Delta E$ and $M_{\rm bc}$ distributions are
different from those of generic $B \bar B$ events. We estimate the background contribution from 
this decay to be 0.1.

%(3) Rare B decay background
A dedicated MC sample comprising of rare $B$ decays is prepared: various decays with 
branching fractions smaller than
$\mathcal{O}({10^{-4}})$ are included and their 
total branching fraction is 
$\mathcal{O}({10^{-3}})$. 
Rare $B$ decays having one or two $K_S^0$ with $\gamma$ in the final state can peak 
in the $M_{\rm bc}$ distribution. 
The backgrounds from the charged $B$ meson pairs do not show any peak in the $\Delta E – M_{\rm bc}$ signal region. 
On the other hand, the background from the neutral $B$ meson pairs peaks in the signal region and 
the largest contribution (34\%) to the peak comes from $B^0 \rightarrow X_{d \bar{d}} (\rightarrow K_S^0 K_S^0 ) \gamma$. 
Herein, $X_{d\bar{d}}$ is a meson composed of a $d{\--}\bar d$ quark pair. 
We regard this as 
 signal because the quark level transition and the final state are the same as for the signal. 
When we treat this decay mode as a signal by using MC information, the peaking background is removed. 
Neutral and charged rare $B$ backgrounds are estimated to be 1.0 and 0.9 events in the signal region,
respectively.

%(4) Other b --> q decay background
Four more rare decay modes which are not included in the rare $B$ MC samples are considered: 
$\mathcal{B}(B^0 \rightarrow K_S^0 K_S^0 \pi^0 ) < 9 \times 10^{-7}$~\cite{KsKspi}; $\mathcal{B}(B^0 \rightarrow K_S^0 K_S^0 \eta) <1.0 \times 10^{-6} $; $\mathcal{B}(B^0 \rightarrow K_S^0 \pi^+ \pi^- \gamma )=1.99\times 10^{-5} $~\cite{K0pipigamma}; $\mathcal{B}(B^0 \rightarrow \pi^+ \pi^- \pi^+ \pi^- \pi^0 )<9.1\times 10^{-3}$~\cite{PDG2020}. 
The first two decay modes occur via a $b\rightarrow s$ quark transition and become background when $\pi^0$ or $\eta$ are replaced by a photon.
$B^0 \rightarrow K_S^0 \pi^+ \pi^- \gamma$ decays occur through a $b\rightarrow s\gamma$ quark transition 
and can be mis-identified as the signal. 
$B^0 \rightarrow \pi^+ \pi^- \pi^+ \pi^- \pi^0$ occurs in the tree level with $b\rightarrow u$ and can be mis-identified as the signal when $\pi^0$ is replaced by a photon. 
We estimate that the background contribution from these four decay modes is negligible.

We estimate the total number of background events in the signal region to be 4.5$\pm$0.7 via the counting method.
To estimate the background events in the signal region 
using an extended unbinned maximum-likelihood
fitting method, we fit the $M_{\rm bc}$ distribution
satisfying 
$-0.16~ {\rm GeV} < \Delta E < 0.09 ~{\rm GeV}$ 
with an ARGUS function ~\cite{ARGUS} and a Crystal Ball line
shape for the continuum and peaking background, respectively.
The fitting parameters of the Crystal Ball line shape are fixed to those for the signal MC.
We obtain 5.6$\pm$0.8 background events in the signal region.
This result is consistent with that of the counting method.

%%%
%%%From here Branching 
%%%
% Signal Efficiency
The signal efficiency depends on the reconstructed $K_S^0$-pair mass ($M_{K_S^0 K_S^0}$) 
as shown in Table~\ref{tab:consum} and is obtained by
performing
an extended unbinned maximum-likelihood fit
to the $M_{\rm bc}$ distribution satisfying $-0.16~ {\rm GeV} < \Delta E < 0.09 ~{\rm GeV}$ and
$5.2 ~ {\rm GeV/}c^2 < M_{\rm bc} < 5.9 ~ {\rm GeV/}c^2$  
in ten $M_{K_S^0 K_S^0}$ bins of equal sizes between $1~{\rm GeV/}c^2$ and $3~{\rm GeV/}c^2$. 

\begin{table*}[!hbt]
  \caption{Summary of the number of observed events ($N_{\rm obs}$), number of estimated background events
($N_{\rm bkg}$), efficiencies ($\epsilon_S$),
upper limits on the signal yield ($S_{90}$), and branching fraction upper limits (U.L.)
at the 90\% C.L. in each $M_{K_S^0 K_S^0}$ bin for the $B^0 \to K_S^0 K_S^0 \gamma$ decay.}
	 \begin{center}
    \begin{tabular}{|c|c|c|c|c|c|c|} \hline 
mass bin(GeV/$c^2$)&  $\epsilon_S$(\%)  & \it{N}$_{\rm bkg}$ & $\sigma_{\rm sys}$(\%) &\it{N}$_{\rm obs}$ & $S_{90}$ & ${\rm U.L.}(10^{-7}$)\\ \hline
	1.0--1.2   & 3.3    &  0.8$\pm$0.3  & 3.2  & 0  & 1.8  &  0.7  \\ \hline
	1.2--1.4   & 3.0    &  0.9$\pm$0.3  & 3.2  & 3  & 6.5  &  2.8   \\ \hline
	1.4--1.6   & 2.7    &  0.8$\pm$0.3  & 3.2  & 1  & 3.6  &  1.7   \\ \hline
	1.6--1.8   & 2.5    &  0.3$\pm$0.1  & 3.2  & 0  & 2.1  &  1.1   \\ \hline
	1.8--2.0   & 2.3    &  0.8$\pm$0.3  & 3.2  & 2  & 5.1  &  2.9   \\ \hline
	2.0--2.2   & 2.2    &  0.2$\pm$0.1  & 3.2  & 1  & 4.2  &  2.5   \\ \hline
	2.2--2.4   & 2.2    &  0.4$\pm$0.2  & 3.2  & 1  & 3.9  &  2.4   \\ \hline
	2.4--2.6   & 2.2    &  0.2$\pm$0.2  & 3.2  & 0  & 2.2  &  1.3   \\ \hline
	2.6--2.8   & 2.3    &  0.0$\pm$0.0  & 3.2  & 1  & 4.2  &  2.3    \\ \hline
	2.8--3.0   & 2.4    &  0.1$\pm$0.3  & 3.2  & 0  & 2.3  &  1.2     \\ \hline
    \end{tabular}
	\end{center}
  \label{tab:consum}
\end{table*}

%%  Systematics
The systematic uncertainties from the number of produced $B\bar B$ pairs and the $\Upsilon(4S)\to B^0 {\bar B}^0$
branching fraction are 1.4\% and 1.2\%~\cite{PDG2020}, respectively.
The systematic uncertainty in the photon detection efficiency is studied 
using radiative Bhabha events and estimated to be 2.0\%~\cite{KSHORT}.
Using a systematic uncertainty of 0.2\% for $K_S^0$ reconstruction efficiency  
and per track uncertainty in efficiency of 0.4\%~\cite{TRACKING} 
leads to our estimate of 1.4\% for the reconstruction efficiency of two $K_S^0$. 
The systematic uncertainty due to the background suppression using the NN selection 
and $\pi^0 /\eta$ veto is 0.6\%~\cite{KSHORT}.
The signal efficiency depends on $M_{K_S^0 K_S^0}$ and the MC statistical uncertainty in the
efficiency varies between 0.5\% and 0.7\% depending on 
$M_{K_S^0 K_S^0}$. 
The total systematic uncertainty is approximately 3.2\% depending on $M_{K_S^0 K_S^0}$, and 
is summarized in Table \ref{tab:Systmatic}. 
\begin{table}[!hbt]
  \caption{Systematic uncertainties.}
	 \begin{center}
    \begin{tabular}{c|c}
    \hline %\Xhline{3\arrayrulewidth}
		Source &	Uncertainty (\%)					 \\ \hline
		Number of $B\bar{B}$ & 1.4 \\
                Branching fraction of $\Upsilon(4S)\to B^0 \bar{B^0}$ & 1.2 \\
		Photon detection efficiency & 2.0 \\
		Two $K_S^0$ reconstruction & 1.4\\
		NB and $\pi^0/\eta$ veto & 0.6  \\ 
		MC statistics in $M_{K_S^0 K_S^0}$ bin efficiency & 0.5--0.7  \\ \hline
		Total  & 3.2 \\ \hline
		\end{tabular}
		\label{tab:Systmatic}
	\end{center}
\end{table}

\begin{figure}[!htb]
\centering
		\includegraphics[width=.4\textwidth]{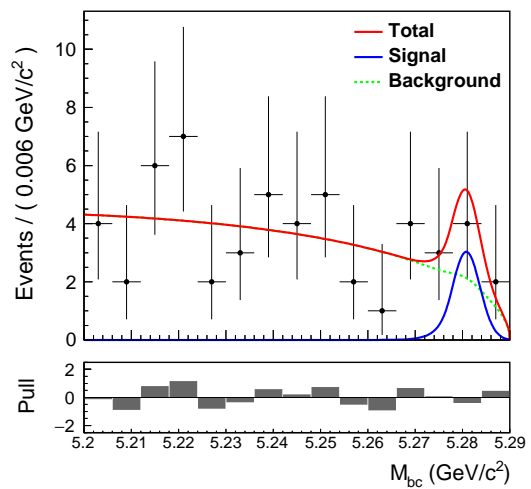}
	\caption{$M_{\rm bc}$ fit distribution with the signal and background parameterized by
         a Crystal Ball line shape and an ARGUS function, 
         respectively.}
	\label{fig:datafit}
\end{figure}

%% upper limit
A total of 9 events 
are observed 
in the signal region. 
As shown in 
Fig.~\ref{fig:datafit}, 
we obtain  
3.8$\pm$3.0 signal and 5.6$\pm0.8$ background events in the signal region 
with an extended unbinned maximum-likelihood fit to
the $M_{\rm bc}$ distribution with a
Crystal Ball line shape including contributions from the peaking background
 for the signal and an ARGUS function for the background, respectively.
The fitting parameters for the signal are fixed to those for the signal MC.
The number of the background events in the signal region agrees well with that of the estimated background events 
in the signal region from MC samples.

\begin{figure}[!htb]
\centering
		\includegraphics[width=.4\textwidth]{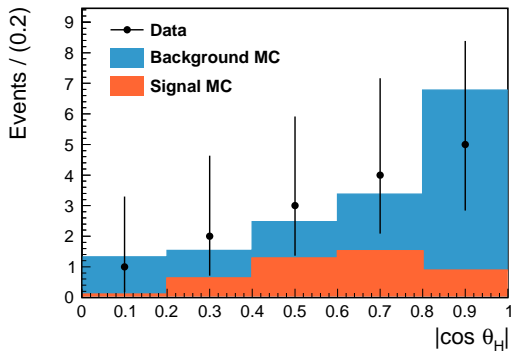}
	\caption{The helicity angle distribution of the observed events in the signal region.}
	\label{fig:helicity}
\end{figure}

The $|\cos\theta_H |$ distributions for events
in the $\Delta E - M_{\rm bc}$ signal region 
are shown in Fig.~\ref{fig:helicity} for data and MC samples.
The signal and background MC samples are normalized to have
the same yields as obtained by fitting the data in the 
signal region. 
The results from data  are consistent with
MC simulation.

\begin{figure}[!htb]
\centering
		\includegraphics[width=.4\textwidth]{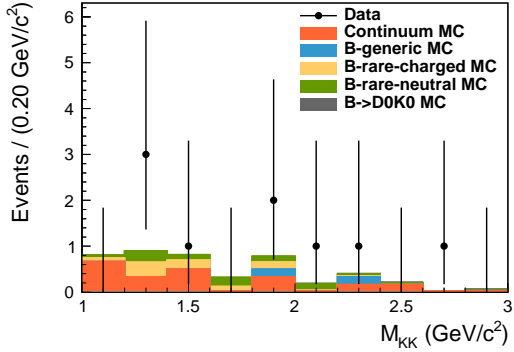} 
	\caption{The $M_{K_S^0 K_S^0}$ distribution in the signal region. The dots represent data and the histograms are the
estimated number of backgrounds from the MC background samples.}
	\label{fig:datamkk}
\end{figure}

The observed number of events in each $M_{K_S^0 K_S^0}$ bin is obtained by counting the events in the 
$\Delta E - M_{\rm bc}$ signal region. 
Figure~\ref{fig:datamkk}  shows the observed number of events ($N_{\rm obs}$) in the full data sample and 
 the estimated background events 
in each $M_{K_S^0 K_S^0}$ bin.
No significant excess over the estimated background is found in data and we derive an upper limit 
for the signal yield ($S_{90}$) at the 90\% confidence level (C.L.) using the POLE program  
by taking into account the uncertainties associated with the 
signal selection efficiency,
background expectation, and systematic uncertainty~\cite{POLE}.
The branching fractions are obtained from 

\begin{equation}
        {\mathcal{B}(B^0 \to K_S^0 K_S^0 \gamma)} = \frac{S_{90}}{\epsilon_S \times N_{B \bar B}},
        \label{eq:s90eq}
\end{equation}

\noindent
where $N_{B \bar B}$ and $\epsilon_S$ are the number of $B \bar B$ pairs and signal efficiency,
respectively.
We set 90\% C.L. upper limits on the partial
branching fractions for the decay $B^0 \to K_S^0 K_S^0 \gamma$
in ten bins of the $K_S^0$-pair for 
1.0 GeV/$c^2 < M_{K_S^0 K_S^0} <$ 3.0 GeV/$c^2$, which
are listed in Table~\ref{tab:consum}.

For the full range 1.0 GeV/$c^2 < M_{K_S^0 K_S^0} <$ 3.0 GeV/$c^2$, we use 
the average efficiency of all bins,  
$(2.5\pm0.4)$\%.
The standard deviation of efficiencies among 
$M_{K_S^0 K_S^0}$ bins is assigned 
as a systematic uncertainty (16.0\%).
Adding to 
other systematic uncertainties listed in Table~\ref{tab:Systmatic}
in quadrature, 
the total systematic uncertainty is 16.2\%.
Using the POLE program with 9 observed events and expected background of $4.5\pm0.7$,
we set the upper limit on  
the branching fraction for the 
1.0 GeV/$c^2 < M_{K_S^0 K_S^0} <$ 3.0 GeV/$c^2$ mass range
to be $5.8\times10^{-7}$ at the 90\% C.L.

We also set upper limits on branching fraction products for
intermediate tensor $f_2$ states, 
$\mathcal{B}(B^0 \to f_2 \gamma)\times \mathcal{B}(f_2 \to K_S^0 K_S^0 )$.
The signal mass regions are taken to be 
1.00 GeV/$c^2 < M_{K_S^0 K_S^0} <$ 1.44 GeV/$c^2$  and
1.44 GeV/$c^2 < M_{K_S^0 K_S^0} <$ 1.63 GeV/$c^2$ for   
$f_2 (1270)$ and  
$f_2^{\prime} (1525)$, respectively.  
These mass regions contain 80\% of signal events.
The results are summarized in Table~\ref{tab:f2reso}.

\noindent 

\begin{table*}[!hbt]
  \caption{Summary of the number of observed events ($N_{\rm obs}$), number of estimated background events
($N_{\rm bkg}$), efficiencies ($\epsilon_S$),
upper limits on the signal yield ($S_{90}$), and branching fraction upper limits (U.L.)
at the 90\% C.L. for the $B^0 \to f_2 \gamma$ and $f_2 \to K_S^0 K_S^0$ decays.}
	 \begin{center}
    \begin{tabular}{|c|c|c|c|c|c|c|} \hline 
 Mode &  $\epsilon_S$(\%)  & $N_{\rm bkg}$ & $\sigma_{\rm sys}$(\%) & $N_{\rm obs}$ & $S_{90}$ & U.L.($10^{-7}$)\\ \hline
	$B^0 \to f_2 (1270)(\to K_S^0 K_S^0 )\gamma$ &  2.3 &1.8$\pm$0.4 & 3.1 & 3  & 5.7  &  3.1    \\ \hline
	$B^0 \to f_{2}^{\prime} (1525)(\to K_S^0 K_S^0 )\gamma$ &  2.2 &0.8$\pm$0.3 & 3.1 & 1  & 3.6  &  2.1    \\ \hline
    \end{tabular}
	\end{center}
  \label{tab:f2reso}
\end{table*}

In summary, we report on the search for radiative $B$ decays with the $K_S^0 K_S^0 \gamma$ final state 
using a data sample of $772 \times 10^6 ~ B\bar B$ pairs. 
No significant signal is
observed for the full data sample.
The signal efficiency depends on $M_{K_S^0 K_S^0}$ and we set upper limits at the
90\% C.L. on the partial branching fractions for the decay
$B^0 \to K_S^0 K_S^0 \gamma$ 
in ten bins of the $K_S^0$-pair 
for 1.0 GeV/$c^2 < M_{K_S^0 K_S^0} <$ 3.0 GeV/$c^2$
to be [0.7--2.9]$\times 10^{-7}$.
We also set an upper limit on its branching fraction
as $5.8\times10^{-7}$ at the 90\% C.L.
for the 1.0 GeV/$c^2 < M_{K_S^0 K_S^0} <$ 3.0 GeV/$c^2$ mass range.
The upper limits at the 90\% C.L. on
the products of  
the branching fractions 
$\mathcal{B}(B^0 \to f_2 \gamma)\times \mathcal{B}(f_2 (1270) \to K_S^0 K_S^0 )$ and
$\mathcal{B}(B^0 \to f_2^{\prime} \gamma)\times \mathcal{B}(f_2^{\prime} (1525) \to K_S^0 K_S^0 )$
are obtained to be $3.1\times 10^{-7}$ and 
$2.1\times 10^{-7}$, respectively. 
These results for the decay $B^0 \to K_S^0 K_S^0 \gamma$
are presented for the first time. 
%%%
%%%From here Acknowledgment 
%%%
% Please paste this acknowledgement into your latex file.
%
% 2019.10.06 updated Russia
\vskip 1mm
  We thank the KEKB group for the excellent operation of the
  accelerator; the KEK cryogenics group for the efficient
  operation of the solenoid; and the KEK computer group, and the Pacific Northwest National
  Laboratory (PNNL) Environmental Molecular Sciences Laboratory (EMSL)
  computing group for strong computing support; and the National
  Institute of Informatics, and Science Information NETwork 5 (SINET5) for
  valuable network support.  We acknowledge support from
  the Ministry of Education, Culture, Sports, Science, and
  Technology (MEXT) of Japan, the Japan Society for the 
  Promotion of Science (JSPS), and the Tau-Lepton Physics 
  Research Center of Nagoya University; 
  the Australian Research Council including grants
  DP180102629, % Sevior
  DP170102389, % Varvell
  DP170102204, % Yabsley
  DP150103061, % Urquijo
  FT130100303; % Urquijo;
  Austrian Federal Ministry of Education, Science and Research (FWF) and
  FWF Austrian Science Fund No.~P~31361-N36;
  the National Natural Science Foundation of China under Contracts
  No.~11435013,  %Zhen-An Liu
  No.~11475187,  %Chang-Zheng Yuan
  No.~11521505,  %Chang-Zheng Yuan
  No.~11575017,  %Cheng-Ping Shen
  No.~11675166,  %Wen-Biao Yan
  No.~11705209;  %Yi-Ming Li
  Key Research Program of Frontier Sciences, Chinese Academy of Sciences (CAS), Grant No.~QYZDJ-SSW-SLH011; % Chang-Zheng Yuan
  the  CAS Center for Excellence in Particle Physics (CCEPP); %Chang-Zheng Yuan,
  the Shanghai Science and Technology Committee (STCSM) under Grant No.~19ZR1403000; %Xiaolong Wang
  the Ministry of Education, Youth and Sports of the Czech
  Republic under Contract No.~LTT17020;
  Horizon 2020 ERC Advanced Grant No.~884719 and ERC Starting Grant No.~947006 ``InterLeptons'' (European Union);
  the Carl Zeiss Foundation, the Deutsche Forschungsgemeinschaft, the
  Excellence Cluster Universe, and the VolkswagenStiftung;
  the Department of Atomic Energy (Project Identification No. RTI 4002) and the Department of Science and Technology of India; 
  the Istituto Nazionale di Fisica Nucleare of Italy; 
  National Research Foundation (NRF) of Korea Grant
  Nos.~2016R1\-D1A1B\-01010135, 2016R1\-D1A1B\-02012900, 2018R1\-A2B\-3003643,
  2018R1\-A6A1A\-06024970, 2019K1\-A3A7A\-09033840,
  2019K1\-A3A7A\-09034974,
  2019R1\-I1A3A\-01058933, 2021R1\-A6A1A\-03043957,
  2021R1\-F1A\-1060423, 2021R1\-F1A\-1064008;
Radiation Science Research Institute, Foreign Large-size Research Facility Application Supporting project, the Global Science Experimental Data Hub Center of the Korea Institute of Science and Technology Information and KREONET/GLORIAD;
the Polish Ministry of Science and Higher Education and 
the National Science Center;
the Ministry of Science and Higher Education of the Russian Federation, Agreement 14.W03.31.0026, % from 15.02.2018
and the HSE University Basic Research Program, Moscow; % from 15.04.2021
University of Tabuk research grants
S-1440-0321, S-0256-1438, and S-0280-1439 (Saudi Arabia);
the Slovenian Research Agency Grant Nos. J1-9124 and P1-0135;
Ikerbasque, Basque Foundation for Science, Spain;
the Swiss National Science Foundation; 
the Ministry of Education and the Ministry of Science and Technology of Taiwan;
and the United States Department of Energy and the National Science Foundation.

%%%
%%%From here Reference 
%%%
%

\end{document}